\documentclass[letterpaper,twocolumn,showpacs,prb]{revtex4}

\usepackage{graphicx}

\def\unit#1{\mathop{\rm #1}\nolimits}%

\begin{document}

\title{Optical transfer cavity stabilization using current-modulated
injection-locked diode lasers}
\author{P. Bohlouli-Zanjani, K. Afrousheh, and J. D. D. Martin}
\affiliation{Department of Physics and Astronomy, University of Waterloo,
Waterloo, Ontario, Canada, N2L 3G1}
\date{June 7th, 2006}

\begin{abstract}
It is demonstrated that RF current modulation of a frequency stabilized
injection-locked diode laser allows the stabilization of an optical cavity
to adjustable lengths, by variation of the RF frequency. This transfer
cavity may be used to stabilize another laser at an arbitrary wavelength, in
the absence of atomic or molecular transitions suitable for stabilization.
Implementation involves equipment and techniques commonly used in laser
cooling and trapping laboratories, and does not require electro- or
acousto-optic modulators. With this technique we stabilize a transfer
cavity using a RF current-modulated diode laser which is injection locked
to a $780\unit{nm}$ reference diode laser. The reference laser is stabilized
using polarization spectroscopy in a Rb cell. A Ti:sapphire ring laser at $%
960\unit{nm}$ is locked to this transfer cavity and may be precisely scanned
by varying the RF modulation frequency. We demonstrate the suitability of
this system for the excitation of laser cooled Rb atoms to Rydberg states.
\end{abstract}

\maketitle

\section{Introduction}

It is often necessary to stabilize lasers at frequencies where direct
locking to an atomic or molecular reference line is not possible. 
Several methods are commonly used for this purpose. One is to stabilize
the target laser by comparison with a second laser (reference laser), which
is stabilized to an absolute frequency reference such as an atomic or a
molecular absorption line. If the frequency of the reference laser is
sufficiently close to the frequency of the target laser, the two lasers may
be heterodyned on a photodetector and the resulting beat note can be used to
stabilize the target laser.\cite{1969-2} However, this is only practical up
to a certain frequency difference due to the bandwidth of the photodetector.

An alternative technique is to use an optical cavity to ``transfer'' the
stability from a stabilized reference laser to the target 
laser.\cite{1991,1998,2002,1979,1994,1996,1982,grabowski2005} 
One
way this may be accomplished is by repetitively scanning the length of a
``transfer cavity'' (TC) with piezoelectric transducers (PZTs). The
transmission fringe positions of the target laser are then compared to the
reference laser using specialized digital circuitry\cite{1991} 
or computers.\cite{1998,2002}
This comparison is used to derive an error signal which may be fed
back to the target laser for stabilization. Using this technique, 
Zhao {\it et al.}~\cite{1998} 
have demonstrated a long-term frequency drift on the order
of $1\unit{MHz}$. However, the scanning rate of the cavity puts a limit on
the maximum rate of error correction. We have also found that this approach
is sensitive to low frequency vibrations. The complexity of the fringe
comparison is an additional drawback.

Scanning the cavity length may be avoided by making the transmission maxima
of both the reference laser and the target laser coincide at the same cavity
length.\cite{1979,1994,1996,1982,grabowski2005}
In this case, the cavity is locked to the reference laser and
the target laser is locked to the cavity using analog circuitry. To make the
fringes coincide it is possible to frequency shift either the reference or
the target laser using an electro-optic modulator (EOM)\cite{1979} or an
acousto-optic modulator (AOM).\cite{1996,grabowski2005} 
In this way, frequency stability and
precise RF frequency tunability can be obtained. Since, in general,
frequency shifts on the order of the free spectral range of the cavity may
be required, the modulator in these systems should be capable of producing a
broad-band of frequency shifts in order to avoid an inconveniently long
transfer cavity.

In this paper, we present a technique for obtaining these frequency shifts
that is inherently broadband and relatively easy to implement without using
AOMs or EOMs. A Fabry--Perot TC is stabilized using a tunable sideband from
a current-modulated diode laser. The carrier of this slave laser is
injection locked to a second diode laser (master laser)\cite{2001} that is
stabilized using an atomic reference. By adjusting the RF frequency of the
current modulation of the injection locked slave laser, the TC may be tuned
to the desired length for stabilization.

\section{Experimental Setup}

The stabilization scheme has been 
developed for a$\ 960\unit{nm}$ commercial ring
Ti:sapphire laser (Coherent MBR-110). This laser is frequency doubled in an
external ring resonator (Coherent MBD-200) to produce approximately $70\unit{%
mW}$ at $\approx 480\unit{nm}$, 
and used with $780\unit{nm}$ lasers to
excite cold Rb atoms to Rydberg states.\cite{pra2006} Since frequency
stabilized $780\unit{nm}$ lasers are required\ in our experiment to laser
cool Rb atoms, they are  convenient references for transfer cavity
stabilization.

The locking procedure is as follows: The $960\unit{nm}$ laser is tuned by
hand to the desired frequency. The master (reference) $780\unit{nm}$ laser
is locked using polarization spectroscopy.\cite{pearman2002} With the TC in
scanning mode, injection locked operation of the slave laser is verified. RF
current modulation is then applied to the slave laser, which produces two
significant sidebands at $\pm f_{m\text{ }}$, the modulation 
frequency.\cite{2001}
The RF modulation frequency is chosen so that a transmission peak of
a $780\unit{nm}$ sideband coincides with a $960\unit{nm}$ transmission peak.
Cavity ramping is then stopped and the cavity is locked to the$\ 780\unit{nm}
$ sideband transmission peak and the $960\unit{nm}$ laser is locked to its
own transmission peak. The lock point of the$\ 960\unit{nm}$ may be varied
by changing the frequency of the slave laser current modulation, $f_{m}.$

\begin{figure}
\includegraphics{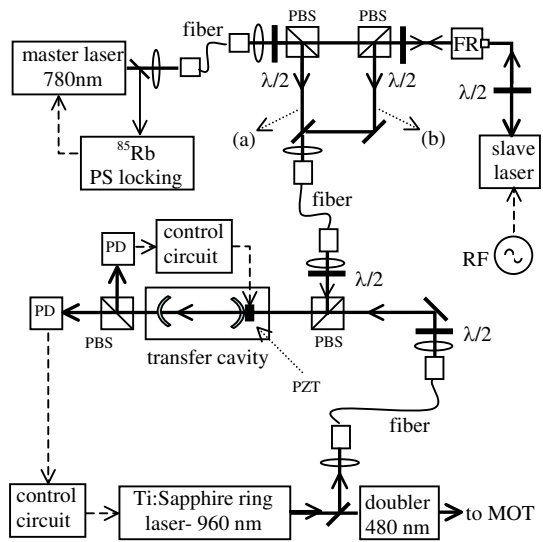}
\caption{\label{f1}
Experimental setup. PBS: polarizing beam splitter; PD:
photodiode; FR: Faraday rotator; PS: polarization spectroscopy.
}
\end{figure}

Figure \ref{f1} illustrates the experimental setup. The reference laser is
an external cavity, grating stabilized, diode laser (Toptica, DL100)
operating at $780\unit{nm}$ with a maximum output power of 
$\approx 150\unit{mW}
$ and short term frequency stability of $\approx 1\unit{MHz}$. The laser can be
coarsely tuned by manually adjusting the grating angle, and fine-tuning is
obtained using a PZT. A small fraction $(10\%)$
of the linearly polarized laser beam
from this laser 
is diverted and divided into two beams. One is used
for Rb saturation absorption spectroscopy (SAS),\cite{preston1996} which
serves as a frequency identifier, and the other beam is directed to a Rb
polarization spectroscopy setup for frequency locking.\cite{pearman2002}
The rest of the reference laser output beam is coupled into a single mode
fiber and the collimated output beam from this fiber is 
used to injection lock the slave laser.

The slave laser is a commercial $780\unit{nm}$ diode laser (Sanyo,
DL7140-201S) in a temperature stabilized mount (Thorlabs, TCLDM9). Current
modulation is applied using a bias-T, driven by a RF
synthesizer (typically operated at 100-400 MHz, 18 dBm). The slave laser is
coupled into a single mode fiber [path (b) in Fig. \ref{f1}]. The output
beam from the fiber is passed through a half wave plate and a polarizing
beam splitter (PBS) to obtain a well-defined polarization. 
This beam is then fed into the TC, which may be temporarily operated 
as a scanning Fabry-Perot
interferometer. A small fraction of the master laser beam is also fed into
the TC for verifying the injection locking [path(a) in Fig. \ref{f1}].

Why not simply current-modulate the reference laser directly\cite{myatt1993}
and lock the cavity to one of the resulting sidebands? Although this will
work in principle, there are a number of reasons to prefer the use of a
current-modulated slave laser, despite the additional complexity. If the
reference laser were directly modulated, the stability of its frequency lock
would be compromised, particularly if the modulation frequency $f_{m}$ were
varied. In addition, by leaving the stabilized laser unmodulated, we can use
part of its beam for other purposes. For example, in our experiment it is
also used as a reference for beat-note locking\cite{1969-2,schunemann1999}
the diode laser
systems used for a magneto-optical trap.

The TC is of the confocal Fabry-Perot type, consisting of two mirrors with
radii of curvature $R=9.18\unit{cm}$, separated by $L \approx  R$ with a
free spectral range of $817\unit{MHz}$. It is desirable to have good
finesse, thus good reflectivity at both $780\unit{nm}$ and $960\unit{nm}$.
It was possible to choose a standard dielectric mirror coating which
exhibits high reflectivity at these two wavelengths (Newport, BD2). A PZT is
mounted on one of the end mirrors. The typical finesse of the TC is $100$
and is limited by beam alignment and difficulty in obtaining the exact
confocal condition.

A small fraction of the Ti:sapphire target laser
beam is coupled into a single mode fiber. On emerging from
the fiber, the light is passed through a PBS to ensure the beam is linearly
polarized, and then aligned into the TC. After the orthogonally polarized $780%
\unit{nm}$ and $960\unit{nm}$ laser beams emerge from the TC, they are
separated by a PBS and directed onto photodiodes.

The TC length is dithered slightly at $1.6\unit{kHz}$ using the PZT by an
amplitude on the order of the cavity linewidth. Lock-in amplifiers are used
to demodulate the transmission through the TC for both 
wavelengths.\cite{1994,1982}
This provides a derivative-like lineshape error signal for locking the
transmission maxima. The $780\unit{nm}$ error signal is used in an
integrator feedback loop which adjusts cavity length using the PZT. This
stabilizes the TC length. The$\ 960\unit{nm}$ error signal is fed into
another integrator control loop which uses the ``Ext Lock'' of the 
target laser control box 
to adjust the frequency. This ``Ext Lock'' control has a
relatively low bandwidth 
( $f_{3\unit{dB}} \approx 10\unit{Hz}$).\cite{mbr} However, this
laser system is pre-stablized using a low-finesse cavity in a similar manner
to the system described in Ref.~\onlinecite{1982}.

\section{Result}

The tuning accuracy and the drift behavior of the frequency locked target
laser is characterized using Rydberg atom excitation in a $^{85}$Rb
magneto-optical trap (MOT). 
The details of this apparatus appear 
elsewhere.\cite{prl2004,pra2006}

The excitation of cold $^{85}$Rb atoms to $46d$ Rydberg states occurs as a
two-color process with nearly resonant $780\unit{nm}$ light ($5s_{1/2}$ - $%
5p_{3/2})$ and $480\unit{nm}$ light ($5p_{3/2}$ - $46d).$ The $780\unit{nm}$
light is necessary for the $^{85}$Rb MOT and is detuned $12\unit{MHz}$ to
the red of the $5s_{1/2}$ F=3 to $5p_{3/2}$ F=4 transition. The $480\unit{nm}
$ light is obtained by frequency doubling the output of a $960\unit{nm}$
Ti:sapphire ring laser -- the target laser for our stabilization scheme.

The $780\unit{nm}$ cooling and trapping light remains on continuously. The $%
480\unit{nm}$ light is pulsed on for $1\unit{%
\mu%
s}$ using an acousto-optic modulator. A pulsed electric field is then
applied to field-ionize any Rydberg atoms and draw the resulting ions to a
microchannel plate detector (MCP). A boxcar integrator is used to gate on
the signal. The excitation and detection sequence repeat at $10\unit{Hz}.$

\begin{figure}
\includegraphics{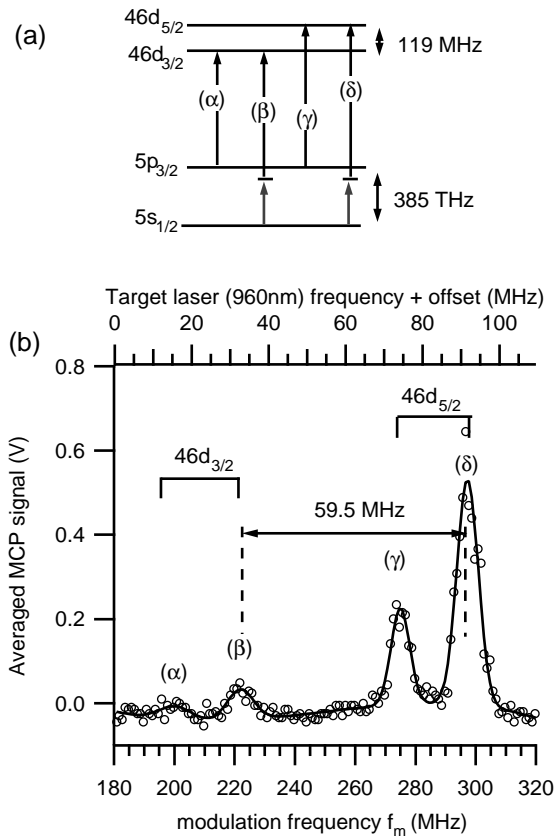}
\caption{\label{f2}
(a) Energy level diagram (b) Spectrum of $^{85}$Rb 
$5p_{3/2}$ to $46d_{3/5}$ and $46d_{5/2}$ Rydberg state transitions obtained
by scanning the RF modulation frequency $f_{m}$. The 
upper horizontal axis is
obtained from Eq. \ref{e0}. The observed peaks correspond to the labelled
transitions shown in part (a). Autler-Townes splitting of the transitions
is observed due to the presence of $780\unit{nm}$ cooling 
laser.\cite{autler1955,teo2003}
With the red-detuning of the $780\unit{nm}$
light (for MOT operation), the ($\beta )$ and ($\delta )$ peaks may be
roughly understood as corresponding to
2-photon absorption from the $5s_{1/2}$
ground state, whereas the ($\alpha $) and ($\gamma $) peaks arise from
step-wise excitation through the $5p_{3/2}$ state.\cite{atomphoton}
}
\end{figure}

When the $960\unit{nm}$ target laser is locked using the scheme described in
the previous section, its output frequency may be scanned by varying the RF
modulation frequency $f_{m}$. Figure \ref{f2} shows the resulting spectrum
in the range of the $^{85}$Rb $5p_{3/2}$ - $46d_{3/2}$ and $5p_{3/2}$ - $%
46d_{5/2}$ transitions. The strong $780\unit{nm}$ field is responsible for
the splitting of the lines into doublets.
This is the  Autler-Townes effect,\cite{autler1955,atomphoton}
similar to the results presented in Ref.~\onlinecite{teo2003}.

We expect the target laser frequency shift $\Delta f_{t}$ to be related to
the slave laser modulation frequency shift $\Delta f_{m}$ by%
\begin{equation}
\Delta f_{t}=\frac{\lambda _{r,air}}{\lambda _{t,air}}\Delta f_{m},
\label{e0}
\end{equation}%
where $\lambda _{r,air}$ and $\lambda _{t,air}$ are the air 
wavelengths of
the reference laser and the target laser. 
In our case, the frequencies of the
reference and target lasers are well-known, but we must estimate the
corresponding refractive indices to determine the air wavelengths. Equation %
(\ref{e0}) can be tested using the observed separation 
of the ($\beta $) and ($\delta $) 
peaks in Fig. \ref{f2}, together with the known $46d_{3/2}$ - $%
46d_{5/2}$ energy separation.\cite{wenhui2003} As shown in Fig. \ref{f2},
the Autler-Townes splitting of $46d_{3/2}$ and $46d_{5/2}$ lines are
identical and thus we do not expect these to contribute to the separation of
the ($\beta $) and ($\delta $) peaks. We find $\Delta f_{t}/\Delta
f_{m}=(0.80\pm 0.015)$, compared to Eq. (\ref{e0}), which predicts 
$\Delta f_{t}/\Delta f_{m}=0.812$. 

\begin{figure}
\includegraphics{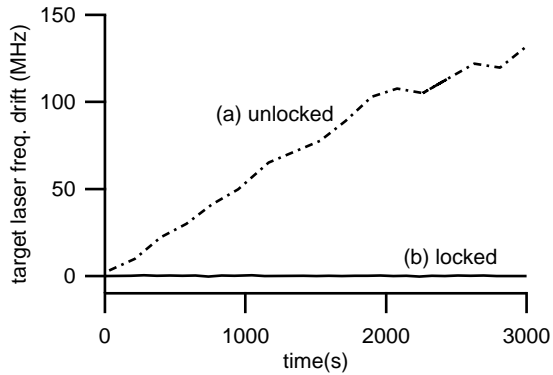}
\caption{\label{f3}
Frequency drift of the target laser system 
(Ti:sapphire, $960\unit{nm}$) 
as a function of time under (a)
unlocked and (b) locked conditions.
}
\end{figure}

By repetitively scanning over the spectrum shown in Fig. \ref{f2} and
recording the peak positions, we can monitor the frequency drift of the
locked target laser. The positions of the ($\alpha $) and ($\gamma $) peaks
are less dependent on the 
frequency fluctuations of the $780 \unit{nm}$ 
cooling laser than
the ($\beta $) and ($\delta $) peaks. Therefore, the stronger ($\gamma $)
line is used to quantify the stability of the target laser. With the locking
system activated, the control voltage applied to the Ti:sapphire laser
varies as time progresses. 
Since the approximate relationship between a
change in the control voltage and the corresponding change in the output
frequency is known, we can use this to estimate the frequency drift that
would have occurred if the laser were not stabilized. Figure \ref{f3} is a
comparison between 
(a) the estimated unlocked and 
(b) the locked frequency drifts
over $\approx 1\unit{hr}$. There is a dramatic reduction in the long-term
frequency drift when the laser is locked.

\section{Performance limitations}

Since the TC is not evacuated, it is limited in performance by
variations in the refractive indices of air for the
target laser and the reference laser wavelengths, $n_{t}$ and $n_{r}$
respectively. 
The environmental influences on the locked target laser frequency 
can be approximated using:
\begin{equation}
\frac{\partial f_{t}}{\partial \alpha }=
\left[\frac{(\partial n_{r}/\partial
\alpha )\text{ }n_{t}-(\partial n_{t}/\partial \alpha )\text{ }n_{r}}{%
(n_{t})^{2}} 
\right]
\text{ }\frac{n_{t}}{n_{r}}\text{ }f_{t},  \label{e1}
\end{equation}%
where $f_{t}$ is the target laser frequency and $\alpha $ represents an
environmental parameter such as pressure, temperature, or humidity. The
resulting sensitivities are tabulated in Table \ref{t1}.

\begin{table}[b] \centering%
\caption{Frequency sensitivity of the locked target laser to environmental
conditions for 
$\lambda_{\rm t, vac} = 960 \: {\rm nm}$,
$\lambda_{\rm r, vac} = 780 \: {\rm nm}$,
$P = 760 \: {\rm torr}$,
$T = 20 \: {\rm ^{\circ}C}$,
${\rm RH} = 50 \%$ and
${\rm CO_{2}} = 450 \: {\rm ppm}$.
To evaluate  Eq. (2), we used the NIST refractive index calculation 
program,\cite{nist} which is based on the Ciddor equation.\cite{ciddor1996}
\label{t1}}%
\begin{tabular}{ll}
\hline\hline
$\mathbf{\alpha }$ & $\ \ \ \ \ \partial f_{t}/\partial \alpha $ \\ \hline
Pressure & $\ \ \ \ 350 \unit{kHz}/\unit{torr}$ \\ 
Temperature & $\ -850 \unit{kHz}/\unit{%
{{}^\circ}%
}$C \\ 
Relative Humidity & $\ \ \ \ 19 \unit{kHz}/$ \% \\ 
\hline\hline
\end{tabular}%
\end{table}%

\begin{figure}
\includegraphics{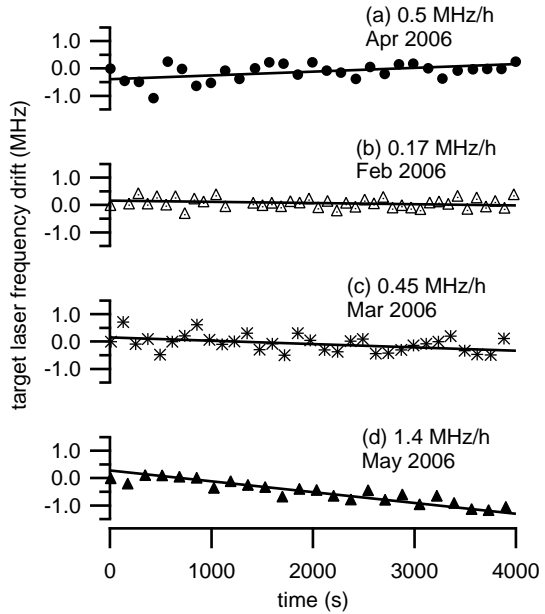}
\caption{\label{f4}
Frequency drift of the locked target laser system
(Ti:sapphire, $960\unit{nm}$) 
for several time periods
over a few months. 
}
\end{figure}

Figure \ref{f4} illustrates the frequency drift of
the locked target laser as a function of time collected at various time over
several months. 
From nine such data sets we observed an average long term ($%
\approx 1\unit{hr}$) frequency drift of $-0.141\pm 0.90\unit{MHz}/\unit{hr}$%
. This is consistent with the typical variation of environmental parameters
listed in Table \ref{t1}. Thus, we expect that the frequency stability of 
the target laser will be improved if the TC is evacuated to minimize the
environmental effects.

Ultimate long-term stability is also limited by the frequency drift of the
reference laser. The reference laser is frequency stabilized using
polarization spectroscopy (PS) in\ a Rb vapor cell.\cite{pearman2002} To
observe the drift of this laser we have monitored the beat note between this
laser and a $780\unit{nm}$ laser stabilized using saturated absorption
spectroscopy with third-harmonic lock-in detection. The relative drift of
these two systems was typically less than $100\unit{kHz}$/hr. 
We have found
polarization spectroscopy locking to be a good compromise between several
factors, including long-term stability, robustness and complexity. 
However, if necessary, less long-term reference laser drift could be 
obtained using alternative techniques.\cite{ye1996,zhu1997,bruner1998}

It is essential to be able to vary the TC length over several free spectral
ranges.  This is to
ensure that the slave laser sideband and $960\unit{nm}$
transmission peaks are well removed from the carrier transmission, which is
stronger and may interfere with cavity locking to the sideband. This
requires a long-extension PZT, which limits the bandwidth of the cavity lock
and consequently the target laser lock. An improved system could use a fast
short extension PZT on one end-mirror and a slower long extension PZT on the
other end-mirror. The fast PZT would be used for dithering and fast cavity
stabilization, whereas the slow PZT would handle long-term 
drift.\cite{hall2000}
With these improvements it is expected that the bandwidth of the
error signal would be sufficient to directly stabilize 
external cavity diode lasers for many applications.

\section{Conclusion}

In this paper, we report a general technique for laser frequency
stabilization at arbitrary wavelengths using a reference laser and transfer
cavity.  A target laser frequency drift of $<1\unit{MHz}/\unit{hr}$ 
has been demonstrated.
The equipment involved is commonly used in laser cooling and
trapping laboratories, and does not require special modulators and drivers.
A controllable frequency source is required, but this is the same as for
electro- or acousto-optic modulators. If
precise RF scanning is not required, the RF synthesizer could be replaced by
inexpensive voltage controlled oscillators, as in Ref.~\onlinecite{2001}. 

\section{Acknowledgements}

It is a pleasure to acknowledge discussions with A. Madej (NRC, Ottawa)
and J. Petrus (Waterloo). We thank M. Fedorov for fabrication and 
testing of the transfer cavity. 
This work was supported by NSERC, CFI, and OIT.


\newpage

\begin{thebibliography}{99}
\bibitem{1969-2} R. L. Barger and J. L. Hall, 
Phys. Rev. Lett. \textbf{22}, 4 (1969).

\bibitem{1991} B. G. Lindsay, K. A. Smith, and F. B. Dunning,
Rev. Sci. Instrum. \textbf{62}, 1656 (1991).

\bibitem{1998} W. Z. Zhao, J. E. Simsarian, L. A. Orozco, and G. D. Sprouse,
Rev. Sci. Instrum. \textbf{69}, 3737 (1998).

\bibitem{2002} A. Rossi, V. Biancalana, B. Mai, and L. Tomassetti,
Rev. Sci. Instrum. \textbf{73}, 2544 (2002).

\bibitem{1979} B. Burghardt, W. Jitschin, and G. Meisel, 
App. Phys. \textbf{20}, 141 (1979).

\bibitem{1994} E. Riedle, S. H. Ashworth, {J. T. Farrell, Jr.}, and D. J.
Nesbitt, Rev. Sci. Instrum. \textbf{65}, 42 (1994).

\bibitem{1996} D. F. Plusquellic, O. Votava, and D. J. Nesbitt,
Appl. Opt. \textbf{35}, 1464 (1996).

\bibitem{1982} J. Helmcke, S. A. Lee, and J. L. Hall, 
Appl. Opt. \textbf{21}, 1686 (1982).

\bibitem{grabowski2005} A. Grabowski, R. Heidemann, R. L\"{o}w, J. Stuhler,
and T. Pfau, 
arXiv, quant-ph/0508082.

\bibitem{2001} R. Kowalski, S. Root, S. D. Gensemer, and P. L. Gould, 
Rev. Sci. Instrum. \textbf{72}, 2532 (2001).

\bibitem{pra2006} K. Afrousheh, P. Bohlouli-Zanjani, J. D. Carter, A.
Mugford, and J. D. D. Martin, 
Phys. Rev. A \textbf{73}, 063403 (2006).

\bibitem{pearman2002} C. P. Pearman, C. S. Adams, S. G. Cox, P. F. Griffin,
D. A. Smith, and I. G. Hughes, J. Phys. B: At.
Mol. Opt. Phys. \textbf{35}, 5141 (2002).

\bibitem{preston1996} D. W. Preston, 
Am. J. Phys. \textbf{64}, 1432 (1996).

\bibitem{myatt1993} C. J. Myatt, N. R. Newbury, and C. E. Wieman, 
Opt. Lett. \textbf{18}, 649 (1993).

\bibitem{schunemann1999} U. Sch\"{u}nemann, H. Engler, R. Grimm, 
M. Weidem\"{u}ller, and M Zielonkowski, 
Rev. Sci. Instrum. \textbf{70}, 242 (1999).

\bibitem{mbr} Operator's Manual Model MBR-110 Single Frequency Ti:Sapphire
Laser, Coherent (2002).

\bibitem{prl2004} K. Afrousheh, P. Bohlouli-Zanjani, D. Vagale, A. Mugford,
M. Fedorov, and J. D. D. Martin, Phys. Rev.
Lett. \textbf{93}, 233001 (2004).

\bibitem{autler1955} S. H. Autler and C. H. Townes, 
Phys. Rev. \textbf{100}, 703 (1955).

\bibitem{atomphoton} C. Cohen-Tannoudji, J. Dupont-Roc, and G. Grynberg
``Atom-Photon Interactions : Basic Processes and Applications'', 
J. Wiley \& Sons, New York,
1998.

\bibitem{teo2003} B. K. Teo, D. Feldbaum, T. Cubel, J. R. Guest, P. R.
Berman, and G. Raithel, 
Phys. Rev. A. \textbf{68}, 053407 (2003).

\bibitem{wenhui2003} W. Li, I. Mourachko, M. W. Noel, and T.F.
Gallagher, 
Phys. Rev. A. \textbf{67}, 052502 (2003).

\bibitem{nist} National Institute of Standards and Technology (NIST)
(http://emtoolbox.nist.gov/Wavelength/Ciddor.asp), 30 March 2006.

\bibitem{ciddor1996} P. E. Ciddor, 
Appl. Opt. \textbf{35}, 1566 (1996).

\bibitem{ye1996} J. Ye., S. Swartz, P. Jungner, and J. L. Hall, 
Opt. Lett. \textbf{21}, 1280 (1996).

\bibitem{zhu1997} M. Zhu and R. W. Standridge, Jr.,
Opt. Lett. \textbf{22}, 730 (1997)

\bibitem{bruner1998} A. Bruner, V. Mahal, I. Kiryuschev, A. Arie, M. A.
Arbore, and M. M. Fejer, 
Appl. Opt. \textbf{37}, 6410 (1998).

\bibitem{hall2000} J. L. Hall, M. S. Taubman, and J. Ye, ``Laser
stabilization'' in Handbook of Optics IV, M. Bass, J. M. Enoch, E. Van
Stryland, and W. L. Wolfe, Eds., Optical Society of America, Washington D.C.,
Chapter 27, McGraw-Hill, New York (2000).
\end{thebibliography}
\end{document}